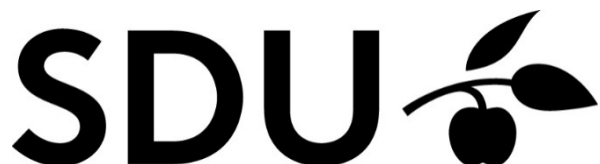

# Energy Flexibility Potential in the Brewery Sector: A Multi-agent based simulation of 239 Danish Breweries

Daniel Anthony Howard, Zheng Ma, Jacob Alstrup Engvang, Morten Hagenau, Kathrine Lau Jørgensen, Jonas Fausing Olesen, Bo Nørregaard Jørgensen

Maersk Mc-Kinney Moller Institute
University of Southern Denmark
Odense, Denmark

*Abstract*— The beverage industry is a typical food processing industry, accounts for significant energy consumption, and has flexible demands. However, the deployment of energy flexibility in the beverage industry is complex and challenging. Furthermore, activation of energy flexibility from the whole brewery industry is necessary to ensure grid stability. Therefore, this paper assesses the energy flexibility potential of Denmark's brewery sector based on a multi-agent-based simulation. 239 individual brewery facilities are simulated, and each facility, as an agent, can interact with the energy system market and make decisions based on its underlying parameters and operational restrictions. The results show that the Danish breweries could save 1.56 % of electricity costs annually while maintaining operational security and reducing approximately 1745 tonnes of $CO_2$ emissions. Furthermore, medium-size breweries could obtain higher relative benefits by providing energy flexibility, especially those producing lager and ale. The result also shows that the breweries' relative saving potential is electricity market-dependent.

*Index Terms*--Industry 4.0, Brewery, Energy Flexibility, Demand Response, Agent-based Modeling

## I. Introduction

The grid stability and security of supply are challenged due to the increasing penetration of renewable energy sources in the electricity grid [1]. Furthermore, conventional balancing of the electricity grid through supply-side management is becoming costly, and the capacity required to ensure the security of supply would be inefficient [2]. Demand-side management has seen increasing potential to mitigate the impact of fluctuations in the electricity grid and aid in stabilization by adjusting consumer demand subject to electricity market conditions [3].

Demand side management can be divided based on the load-shape objective, e.g., peak clipping, valley filling, and load shifting [4]. An option for demand-side management that has been proven effective is demand response. Demand response relies on the demand side adjusting its consumption through, e.g., load shifting, and peak clipping based on the current electricity market conditions. Theoretically, demand response can happen in the residential, industrial, and tertiary sectors (trade, commerce, and services) [5]. The industry has a high potential demand response compared to residential and tertiary sectors due to their high electricity consumption and energy management systems [6].

The process industry accounts for a significant portion of energy-intensive industries, e.g., food, pulp and paper, refining, iron and steel, and chemical plants [7]. Therefore, the saving opportunities in process-type facilities will often be greatest in energy efficiency and flexibility measures. As IEA [8] pointed out, there is a need to reduce energy consumption and carbon emissions in energy-intensive industries. However, there is hesitance to adopt energy flexibility measures as there is an underlying uncertainty of the implications on production [8]. Furthermore, the production flow with interconnected steps makes the deployment more complex and challenging [9].

The beverage industry is a typical food processing industry and accounts for significant energy consumption, e.g., 1 % of Danish energy consumption [10]. The beverage industry can be further divided based on the beverage type, with beer production being the category with the highest energy consumption accounting for 40 % of the beverages industry's combined energy consumption [10]. For instance, Denmark has the highest number of breweries per capita [11] among the European nations. As of April 2022, there were 275 breweries in Denmark. A survey based on the Danish Brewery Associations members shows that approximately 50 % of Danish beverage facilities might be permanently close or go bankrupt due to COVID-19 and the increasing energy prices [12].

60 % of the brewery's energy consumption is consumed for refrigeration and packaging [13]. A single brewery could save approximately 9 % on its energy cost through electricity price optimization of the cooling supply [14]. The brewery facilities can be energy flexible and reduce the peak load energy consumption by 32 % [15]. However, it is difficult to utilize the energy flexibility by only focusing on a single facility because altering the operation of a single facility may propagate to the whole process flow [9].

Meanwhile, to ensure grid stability and activate national demand response programs, realizing energy flexibility from one brewery production is not enough, but the whole brewery industry. However, no literature has investigated the demand response potential for the whole beverage industry. Furthermore, there is a need to compare the implications of varying facility types and sizes as they respond differently to demand response strategies. To capture the varying facility types, agent-based modeling is used. Agent-based modeling focuses on the internal logic of individual agents within the system. Through agent interaction and environment state, the system's behavior becomes a result of the combined agent logic, i.e., emergent phenomena. Agent-based modeling enables the development of production process-specific agents contained in the individual brewery, with the population of breweries collectively contained in the modeling of the Danish brewery sector.

Therefore, this paper aims to assess the energy flexibility potential of the whole brewery industry in the case of Denmark. This paper extends the previous work on the brewery fermentation tanks [16] to include the entire production process. Furthermore, this paper investigates the benefits of different brewery production segments' participation in the electricity market. Moreover, the energy flexibility potential of the whole Danish brewery industry is evaluated to understand its contribution to the Danish electricity grid stability.

The rest of this paper is structured as follows. Initially, the structure of the Danish electricity system is introduced together with the generic brewery production process and underlying behaviors. Afterward, the agent population characteristics are introduced in the methodology. The results review the potential for the brewery industry to participate in implicit demand response in varying annual production volume segments. Lastly, the findings are discussed and concluded.

## II. BACKGROUND

### A. Danish Electricity System

The Danish electrical grid is divided into a transmission system, a distribution grid, and consumption. It is split into two connected areas DK1 (West Denmark), which is geographically constituted by Jutland and Funen, and DK2 (East Denmark), which is geographically located in Zealand. The two areas are connected by a high voltage direct current (HVDC) connection. Furthermore, Denmark is connected to neighboring countries through a number of interconnectors to Norway, Sweden, Germany and Netherlands with connection to England currently under construction.

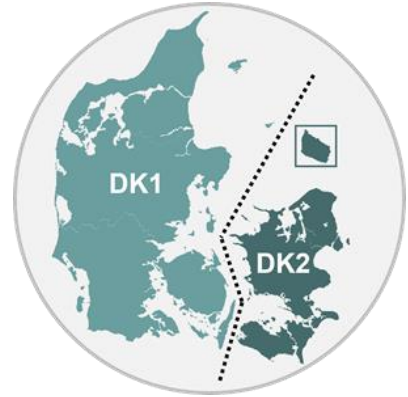

Figure 1. Divison of the Danish synchronous areas [17].

The generation of electricity is inputted into the 400 kV grid. The transmission grid operates on 400 kV and 150 kV for DK1 and 132 for DK2; this part of the electricity system is operated and managed by the transmission system operator (TSO). In Denmark, there is a single TSO, Energinet. The electricity is subsequently stepped down to the distribution grid, which operates at 60 kV and 50 kV for DK1 and DK2, respectively, before being further stepped down to 10 kV and 0.4 kV. The distribution grid is managed by the distribution system operator (DSO). Consumers are connected to different parts of the electricity grid based on their demand for electricity. Demand response is mainly concerned with shifting the timing of consumption as a response to compensation. Demand response can be divided into implicit and explicit approaches. Implicit DR involves consumers responding to a price signal reflecting current market conditions. The viability of implicit DR relies on installing smart meters at the consumer to register the consumption in an appropriate time granularity. Equipped with the knowledge of the price signal, the consumer can implement internal measures for responding. Explicit demand response has higher complexity than implicit demand response and involves multiple parties and requirements for the consumer; however, the monetary benefit of participating in a specific, explicit demand response program is significantly higher than implicit demand response [18]. Different legislative parameters are in force based on the country, setting requirements for consumers wishing to participate in explicit demand response programs in the balancing market. A significant requirement in the Danish system is the minimum capacity and response time required to participate [17].

### B. Brewery Production Process

The brewery production can be described as a series of process steps that convert the raw materials of primarily malted barley, water, and hops to the final beer product through the addition of energy and time. The brewery process is an established practice with little deviation. The primary differences between breweries is observed in the product line. A benchmarking of Danish microbreweries estimated the electrical energy consumption to be in the range of 22 - 106 kWh/hl. A generic brewery production process is shown in Figure 2.

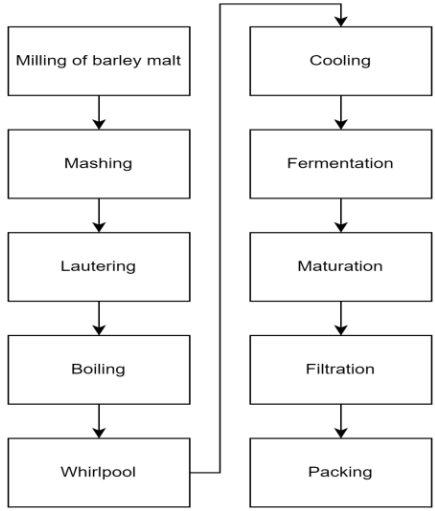

Figure 2.  Brewery production process

The brewery production process is shown in Figure 2. It starts with the milling of barley malt. Some brewery facilities germinate barley alone; however, this was not considered in this study. After milling, the grain is mixed with water in the masher allowing the mix to infuse; the infusion is performed while heating the mixture. After milling, the mash is transferred to the lauter tun to separate the wort (unfermented beer); the lautering is typically done around 77 °C. After lautering, the residual grain is removed, and the wort is transferred to the kettle. In the kettle, hops are often added during the heating in specific stages depending on the beer. The kettle's heating stages depend on the beer and follow the brewery's recipe. After the kettle, the wort is transferred to the whirlpool, which separates any remaining solids from the wort that is subsequently cooled. The wort is transferred to the fermentation tanks, where yeast is added; the wort stays in the fermentation tanks for several days to complete primary fermentation. After primary fermentation, the wort has been transformed into green beer, which is moved to a conditioning vessel for maturation. After conditioning, the beer is ready for tapping. Sometimes, a cold crash is conducted after conditioning to clarify the beer. Subsequently the beer enters the packing stage in which it is prepared for sales.

*1) Fermentation Process*

A significant complexity in the brewery is the fermentation process which has significant underlying uncertainties associated with it. In previous work, the fermentation process is either unaccounted for or lacks temporal development. The fermentation rate is described using the non-linear regression model presented in [19] to model the heat production of fermentable sugars.

$$P_t = P_e + \frac{P_i - P_e}{(1 + e^{-B(t-M)})^{\frac{1}{s}}} \quad (1)$$

Using the function for fermentation heat used in [20], the heat development for a batch of unfermented beer can be described as:

$$Q(t) = \rho_{wort}(°P,T) \cdot V \cdot \frac{d}{dt}P_t(t) \cdot e \quad (2)$$

Where the density and specific heat capacity of the wort is described using the empirical relationships established in [21], the fermentation time of the wort will be influenced by the type of beer, primarily categorized as either lager or ale. As the fermentation process has been identified as a process with significant energy flexibility potential within the brewery, the energy flexibility within this paper will focus on the fermentation tanks [14].

### III. METHODOLOGY

The brewery production process consists of specific steps, as seen in Figure 2. The modeling of the processes can be characterized as discrete events with the wort transitioning from one process to another at specific points in time. However, modeling the fluid's properties and the heat exchange follows system dynamics modeling with continuous development. Accounting for several individual production batches of wort in multiple processes would make the overall brewery system dynamics challenging to represent. Previous studies have shown the ability to use agent-based modeling and simulation to capture complex systems [22]. For simulating the Danish brewery population, AnyLogic was chosen due to its support for multi-method modeling and object-oriented programming. Thereby, the inherent modeling logic can be represented within the individual agents [23].

*A. Danish Brewery Sector Modeling*

As it is unfeasible to collect detailed information production information for all individual Danish breweries, general attributes and characteristics were collected based on publicly available information and previous literature. GIS information on Danish breweries could be extracted from the website [24]. The synchronous area where the brewery is located could be determined using address geocoding based on the respective longitudinal coordinates. Based on the information extraction, 239 breweries could be found for analysis in this paper.

Instantiating the breweries is based on the production capacity of the individual breweries. The yearly production is a result of the operation of the individual brewery. Hence providing production quantity parameters for the individual brewery can be used for inferring the approximate process parameters, e.g., the number of fermentation tanks needed to accommodate the yearly production volume. The approach could also be extended to evaluate other industry segments on a national or international scale through knowledge of segment production statistics.

The consumption and brewery facility setup was based on observed distributions for brewery sizes and adjusted to match the extracted number of Danish breweries and the annual production volume of Danish breweries. The utilized distribution of brewery sizes can be seen in Table I based on [25, 26].

TABLE I. DISTRIBUTION OF DANISH BREWERIES

| Category | Annual Volume [hectoliter] | n | ale:lager | Brewdays/week |
|---|---|---|---|---|
| *Small breweries* | | | | |
| 1 | <680 | 181 | 4:1 | 3 |
| 2 | <5100 | 40 | 4:1 | 3 |
| *Medium breweries* | | | | |
| 3 | <10000 | 6 | 7:3 | 5 |
| 4 | <17500 | 4 | 7:3 | 5 |
| 5 | <36000 | 3 | 7:3 | 5 |
| *Large breweries* | | | | |
| 6 | <70000 | 1 | 3:2 | 7 |
| 7 | < 1350000 | 3 | 3:2 | 7 |
| *Extra large breweries* | | | | |
| 8 | 1350000+ | 1 | 1:1 | 7 |

The ale-to-lager ratios were based on smaller breweries' tendency to focus on ale-based beer due to reduced fermentation time and production expenses [27]. Using the presented ratios for lager and ale production in various size categories of breweries combined with the average number of brewing days per week and working weeks in a year, the number of lager and ale brew cycles could be determined. As the dimensioning of the fermentation tanks is significant for the thermodynamic properties, the fermentation tanks were dimensioned based on the 2:1 ratio for height and diameter. The 2:1 ratio has been established as a suitable aspect ratio for modeling [28]. The tank dimensions could be established through geometrical relationships based on the ratio. The distribution of breweries was done following Table I. Upon initializing the multi-agent system simulation, each brewery is prescribed a random size category. An annual volume was prescribed to the brewery using a triangular distribution using the size category information. The mode of each category's triangular distribution is determined from scaling the mode of the original categories described in [26].

*B. Brewery Agents Simulation Architecture*

Several generic brewery-specific agents were developed to enable the representation of individual breweries in the Danish brewing sector. The internal brewery agents were created by instantiating the individual agents based on the parameters given through the population characteristics described in Table I. The internal and external brewery agents developed for the simulation can be seen in Table II.

TABLE II. DEVELOPED BREWERY-SPECIFIC AGENTS

| Agent name | Functionality | Primary Interaction | N |
|---|---|---|---|
| Denmark | National agent, providing a collective environment for the breweries to reside in. Holds information on the current weather conditions. | n/a | 1 |
| Energy system operator | Note, that two agents were created related to each synchronous area. | n/a | 2 |
| Brewery | Industrial facility agent, the top-level agent which contains all the brewery-specific agents in its environment. | Energy system operator Outdoor weather | 239 |
| Beer | Product agent, contains current temperature and volume of the beer. The beer agent also represents the wort. Contains the yeast development logic | n/a | D[a] |
| Batch | Logisitcal agent, tying the product agent to specific logistical parameters, e.g., start time and deadline. | Planning | D[a] |
| Bottle Filling | Process agent, fills the beer into bottles. Modeled as a sink for completed beer. | Beer | 1 |
| Fermentation Tank | Process agent, contains the beer agent for a specified amount of time. | Beer, Refrigeration unit | D[a] |
| Kettle | Process agent, contains the beer agent and boils it for a specified amount of time. | Beer | 1 |
| Malt | Resource agent, used for creating the beer agent. Holds an amount that dictates the volume of beer. | Beer | D[a] |
| Lautering | Process agent, contains the beer for a specified amount of time. | Beer | 1 |
| Mashing | Process agent, combines the water and malt to create the beer agent that inherits properties. | Beer, Malt | 1 |
| Milling | Process agent, prepares the malt for usage in the masher. | Malt | 1 |
| Whirlpool | Process agent, contains the beer for a specified amount of time | Beer | 1 |
| Refrigeration unit | Provides cooling for the fermentation tanks | Fermentation tank | 1 |

a. Dynamic population size

As seen in Table II, several agents were developed for the individual brewery simulation. The individual brewery was created using a single production line for brewing, with sufficient capacity for handling the volume of beer allotted from the population parameters. As seen in Table II, several types of agents were developed according to the underlying functionality of the agent, e.g., process agent or product agent. The hierarchy of the agents follows the first three agents of Table II. A national agent representing Denmark was created containing weather information and the energy system operator. The breweries are placed in the national agent and connected to the energy system operator associated with their current location, i.e., synchronous area. Thereby, through a common energy system operator interface, the brewery agents can perform flexible operations based on the state of the specific energy system to which it is connected.

IV. RESULTS

Initially, the breweries' current operation without energy flexibility was examined to provide a baseline for comparison. Subsequently, the simulation was conducted to consider implicit demand response. The simulation was run for the entire year 2021.

## A. Danish Brewery Sector

The brewery agents were placed at their respective locations in Denmark using the extracted GIS data seen in Figure 3. The placement of breweries in the simulation corresponds to the brewery's provided geographical address meaning that it could be the location of the company headquarters for large facilities.

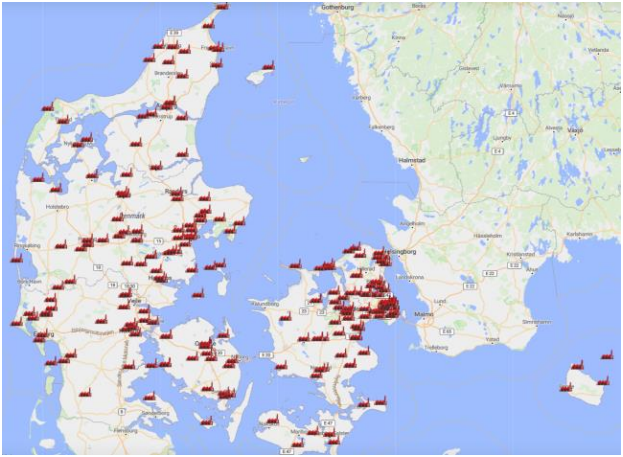

Figure 3. Simulation overview showing Danish brewery locations

Examining the current fermentation tank cooling load demand across the agent populations can be seen in Table III.

TABLE III. BASELINE COOLING LOAD COST FOR DANISH BREWERIES.

| Scenario | Electricity cost [mDKK] | $CO_2$ emissions [tonne] | Load [MWh] |
|---|---|---|---|
| Baseline | 55.45 | 11731.63 | 81990.47 |

Examining the distribution of costs associated with the brewery category from Table III and the synchronous area can be seen in Figure 4.

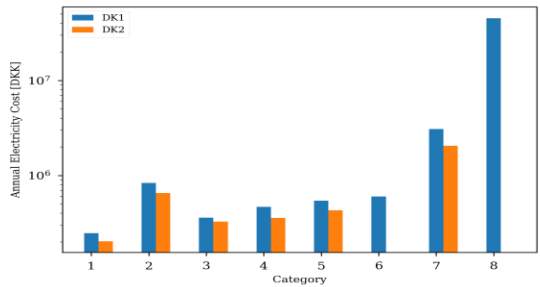

Figure 4. Cooling load cost distribution of brewery categories

As remarked in [29], temperature fermentation temperatures can be altered with minimum impact on the beer's overall taste profile and quality. Therefore, some temperature fluctuation can be allowed in the brewing process, enabling the use of implicit demand response to reduce the overall operation cost. The flexible operation results can be seen in Figure 5.

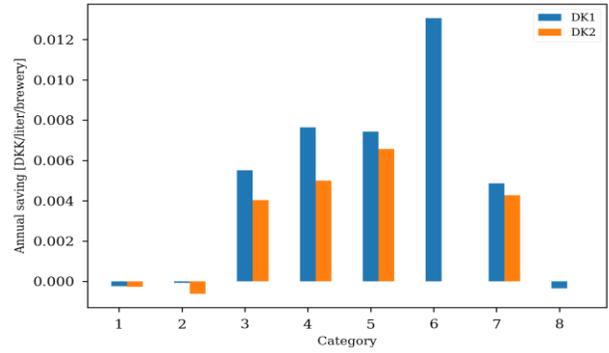

Figure 5. The relative flexibility potential across brewery categories

As seen in Figure 5., category six has the most significant relative flexibility potential. Furthermore, the breweries in category two have the lowest relative potential.

## V. DISCUSSION

The current implementation of the presented solution shows the ability of breweries in the Danish electricity system to benefit from implicit demand response. The solution could be extended to include the bottling and packing part of the brewery, which also contains significant energy-intensive processes. Comparing Figures 4 and 5 with the population distribution presented in Table 1, it is evident that small-scale breweries focusing on producing ale have the less overall potential for partaking in implicit demand response. Furthermore, it should be considered that many small brewers are craft beer breweries and therefore emphasize the correct development and quality of beer. They may be less willing to risk their quality of production. As seen from the relative potential in Figure 5., the relative flexibility potential of medium-large scale breweries is significantly higher compared to the small and extra large. Furthermore, it is evident that the relative saving in the DK1 synchronous area is generally higher compared to the DK2 synchronous area. As it has been established that the breweries can provide implicit demand response, the potential for partaking in explicit demand response could be reviewed in future research.

## VI. CONCLUSION

The process industry contains significant potential for providing implicit demand response that can aid in balancing the electricity system in response to increasing fluctuating input of renewables. However, currently, there is a lack of assessment examing the different consumer types within an industrial sector to determine which consumers can benefit the most. Therefore, this paper presents a case study for the Danish brewery sector, examining the potential for implicit demand response in the fermentation tanks while accounting for the up- and downstream production processes within the brewery. Using multi-agent systems and agent-based modeling enabled the representation of brewery agents within the Danish electricity system. Each brewery contained several process-specific agents representing the brewery production process. The Danish brewery sector was represented using eight categories based on population statistics simulating 239 total breweries. The relative flexibility potential showed that Danish breweries producing between 10000 and 1350000 hectoliters annually had the highest benefit from partaking in implicit demand response, while breweries producing less than 5100 hectoliters annually had little benefit. The benefit could be seen as a response to the ratio of ale and lager produced in specific breweries, as ale dominant breweries showed less overall potential, possibly due to higher fermentation temperature and less volume. The approach presented in this paper can be generalized to other industrial sectors for evaluating national flexibility potentials in industry segments.

ACKNOWLEDGMENT

This paper is funded by the authors' affiliation: SDU Center for Energy Informatics. A special thanks to Bryggeriet Vestfyn for providing an opportunity for visiting and baseline data.